\def\kms{$\mathrm {km}~\mathrm{s}^{-1}$} 
\def\om{\Omega_{\rm p}} 
\def\len{a_B} 
\def\lag{D_L} 
\def\pin{{\cal X}} 
\def\kin{{\cal V}} 
\def\vpd{{\cal R}} 

\documentclass[12pt,preprint]{aastex} % preprint           

\lefthead{Corsini et al.} 
\righthead{The bar pattern speed of NGC 4431} 
      
\begin{document}      
 
\title{The bar pattern speed of the dwarf galaxy NGC 4431$^1$} 
\footnotetext[1]{Based on observations collected at the European  
Southern Observatory, Chile (ESO N$^o$ 71.B-0138(A)  
and N$^o$ 73.B-0091(A)). }

\author{E.~M.~Corsini\altaffilmark{2},      
J.~A.~L.~Aguerri\altaffilmark{3}, 
Victor~P.~Debattista\altaffilmark{4,5}, 
A.~Pizzella\altaffilmark{2}, 
F.~D.~Barazza\altaffilmark{6}, and  
H.~Jerjen\altaffilmark{7}}      
 
\altaffiltext{2}{Dipartimento di Astronomia, Universit\`a di Padova,      
  Vicolo dell'Osservatorio 3, I-35122 Padova, Italy, 
  {\tt enricomaria.corsini@unipd.it},
  {\tt alessandro.pizzella@unipd.it}}
\altaffiltext{3}{Instituto de Astrof\'{\i}sica de Canarias, V\'{\i}a      
  L\'actea s/n, E-38200 La Laguna, Spain, {\tt jalfonso@iac.es}}      
\altaffiltext{4}{Astronomy Department, University of Washington, 
  Box 351580, Seattle, WA 98195-1580, USA, 
  {\tt debattis@astro.washington.edu}}  
\altaffiltext{5}{Brooks Prize Fellow}     
\altaffiltext{6}{Department of Astronomy, University of Texas at Austin,
  1 University Station, C1400, Austin, TX 78712-0259, USA,
  {\tt barazza@astro.as.utexas.edu}}
\altaffiltext{7}{Research School of Astronomy and Astrophysics,  
  Mt. Stromlo Observatory, Australian National University,   
  Cotter Road, Weston ACT 2611, Australia, 
  {\tt jerjen@mso.anu.edu.au}} 
 
\begin{abstract} 
We present surface photometry and stellar kinematics of NGC 4431, a
barred dwarf galaxy in the Virgo cluster undergoing a tidal
interaction with one of its neighbors, NGC 4436. We measured its bar
pattern speed using the Tremaine-Weinberg method, and derived the
ratio of the corotation radius, $\lag$, to the bar semi-major axis,
$\len$. We found $\lag/\len =0.6^{+1.2}_{-0.4}$ at $99\%$ confidence
level. Albeit with large uncertainty, the probability that the
bar ends close to its corotation radius (i.e., $1.0 \leq \lag/\len
\leq 1.4$) is about twice as likely as that the bar is much shorter
than corotation radius (i.e., $\lag/\len > 1.4$).
\end{abstract}


%\date{{\it Draft version on \today}} 
 
\section{Introduction} 
\label{sec:intro} 
 
Stellar bars are observed in a substantial fraction of nearby disk
galaxies (Knapen et al. 2000; Eskridge et al. 2000; Marinova \& Jogee
2006). Their growth is partly regulated by the exchange of angular
momentum with the stellar disk and the dark matter (DM) halo. For this
reason the dynamical evolution of bars can be used to constrain the
distribution of DM in the inner regions of galaxy disks.
Using perturbation theory Weinberg (1985) predicted that a bar would
lose angular momentum to a massive DM halo through dynamical friction,
slowing down in the process.
This prediction was subsequently confirmed in $N$-body simulations
(Debattista \& Sellwood 1998, 2000; Athanassoula 2003; O'Neill
\& Dubinski 2003; Sellwood \& Debattista 2006). They found that
bars are slowed efficiently within a few rotation periods if a
substantial density of DM is present in the region of the bar. On the
other hand, if the mass distribution is dominated by the stellar disk
throughout the inner few disk scale-lengths, then the bar remains
rapidly rotating (here defined as the bar ending just inside its
corotation radius) for a long time. Thus the accurate measurement of
the bar pattern speed, $\om$, provides a way to discriminate whether
the central regions of the host galaxy are dominated by baryons or by
DM.
 
Bar pattern speeds are most usefully parametrized by the
distance-independent ratio $\vpd = \lag / \len$, where $\lag$ is the
corotation radius and $\len$ the semi-major axis of the bar. Bars
which end near corotation ($1 \leq \vpd \leq 1.4$) are termed fast,
while shorter bars ($\vpd > 1.4$) are said to be slow.
The best way to determine the bar rotation parameter $\vpd$ is by
measuring $\om$ with the model-independent method developed by
Tremaine \& Weinberg (1984, hereafter TW). To date this method has
been used to measure $\om$ of mainly massive ($V_c \ga 150$ \kms)
early-type barred galaxies (see Corsini 2004 and references therein)
and it has been successfully tested against $N$-body simulations
(Debattista 2003; O'Neill \& Dubinski 2003). All measured bars are
compatible with being fast, implying that the inner regions of these
galaxies are not dominated by DM.
The only exception appears to be the slow bar in NGC~2915, a blue
compact dwarf galaxy, for which the TW method was applied to
H~{\scriptsize I} radio synthesis data (Bureau et al. 1999).

The minimum possible amount of DM in disk galaxies can be determined
from the rotation curve, by scaling up the mass-to-light ratios of the
luminous components to fit the central velocity gradient (see Bosma
1999 for a review). However, the lack of any obvious transition region
from the bulge/disk-dominated to the halo-dominated part of rotation
curves severely limits the ability to obtain a unique decomposition
and derive the central DM distribution (e.g., Corsini et al. 1999).
The structure of DM halos is more directly revealed in galaxies where
the luminous components are thought to give a nearly zero contribution
to the mass budget, because this makes mass modeling easier and the
derived DM distribution less uncertain. For this reason dwarf galaxies
were considered ideal targets for studying the properties of DM halos
and for testing if they have a central cuspy power-law mass density
distribution as predicted by cold dark matter (CDM) cosmology (Navarro
et al. 1996, 1997; Moore et al. 1998, 1999).
However, recent extensive studies have shown that in most of them the
contribution of the stellar disk can account for the inner rotation
curve and DM halos with a constant-density core provide a better fit
to than those with a density cusp (e.g., de Blok \& Bosma 2002;
Swaters et al. 2003; Spekkens et al. 2005).
 
Nonetheless, the interpretation of these results has been
controversial because of possible observational systematic effects,
such as the poor spatial resolution of radio maps (van den Bosch \&
Swaters 2001) and slit misplacement in optical spectroscopy (Swaters
et al. 2003) as well as non-circular (Hayashi \& Navarro 2006), and
off-plane gas motions (Swaters et al. 2003b; Valenzuela et al. 2005).
 
Since measurements of pattern speeds with the TW method do not require
high spatial resolution or slits passing exactly through the galaxy
center and refers to the stellar component only, such a measurement in
a dwarf disk galaxy can test unambiguously for the presence of a dense
DM cusp.
In the present Letter we attempt such a measurement for NGC~4431.

\section{NGC 4431} 
\label{sec:n4431} 
 
NGC~4431 (VCC 1010) is a small ($1\farcm7\times1\farcm1$ [de
Vaucouleurs et al. 1991, hereafter RC3]) and faint ($B_T=13.74$ [RC3])
nucleated dwarf galaxy (Binggeli et al. 1985) classified as dSB0/a by
Barazza et al. (2002) following an unsharp mask analysis of the
galaxy. The presence of a bar had earlier been missed by virtue of the
galaxy's low luminosity and the relatively small angle between the
bar's and disk's position angle (PA).  Both the presence of a strong
bar and trailing arms (which are clearly visible in Fig. 3 of Barazza
et al. 2002) and the rotationally-supported kinematics (Simien
\& Prugniel 2002; Pedraz et al. 2002) are indicative of a genuine disk
galaxy.  The total absolute magnitude of the galaxy is
$M_{B_T}^0=-17.41$ corrected for inclination and extinction (RC3) and
adopting a distance of 16.2 Mpc (Jerjen et al. 2004).
It is located close to the center of the Virgo cluster at a projected
distance of $\sim300$ kpc from M~87. Its closest neighbor in
projection is the dE6/dS0 galaxy NGC 4436, at a projected separation
of $3\farcm7$ or $\sim18$ kpc.

\section{Broad-band imaging}      
\label{sec:imaging} 
      
Deep broad-band imaging of NGC~4431 was carried out with the Very
Large Telescope (VLT) at the European Southern Observatory on April 1,
2000 as part of a project aimed at measuring distances, metallicities,
and ages of dwarf galaxies in the Virgo cluster. Details of
observations, data reduction, and analysis are reported elsewhere
(Barazza et al. 2003; Jerjen et al. 2004).
 
We analyzed the $R-$band image of NGC~4431 obtained by Barazza et
al. (2003), with foreground stars and background galaxies subtracted.
We fitted ellipses to galaxy isophotes with the {\tt
IRAF}\footnote{{\tt IRAF} is distributed by NOAO, which is operated by
AURA Inc., under contract with the National Science Foundation} task
{\tt ELLIPSE}, after masking the northeastern part of the image to
minimize the light contamination from NGC 4436, the outskirts of which
fall in the field of view of the image. We first fitted ellipses
allowing their centers to vary to test whether patchy dust obscuration
and tidal deformations were present. Having found no evidence of
varying ellipse centers within the errors of the fits for the inner
$40''$ we concluded that in these regions there is little or uniform
obscuration and no tidal deformations, and that the galaxy is a viable
candidate for the TW method. At larger radii, the shift of ellipse
centers ($\approx1''$) of NGC 4431 in the direction of NGC 4436 may be
due to the interaction between the two galaxies. The ellipse fits were
then repeated with the ellipse center fixed and the resulting surface
brightness profile is plotted in Fig. 1d. The inclination
($i=48\fdg7\pm1\fdg5$) and disk PA ($\rm PA_{\it
disk}=6\fdg9\pm1\fdg0$) were determined by averaging the values
measured between $50''$ and $70''$. They are consistent within the
errors with values obtained by fitting ellipses with free centers.
 
We measured $\len$ (Fig. 1) using four independent methods based on
Fourier amplitudes (Aguerri et al. 2000), Fourier and ellipse phases
(Debattista et al. 2002), and a parametric decomposition of the
surface brightness profile (Prieto et al. 2001). The mean of the
resulting values is our best estimate of $\len$ and we assume the
largest deviation from the mean as our error estimate
($\len=21\farcs9\pm1\farcs5$). The bar length is in agreement with the
radius where the spiral arms seen in the unsharp mask image of Barazza
et al. (2002) start.

\section{Long-slit spectroscopy}  
\label{sec:spectroscopy}     
       
The spectroscopic observations of NGC 4431 were carried out in service
mode at VLT on May 4-6, 2003 (run 1), April 21-22, 2004 (run 2) and
May 18-23, 2004 (run 3).
The Focal Reducer Low Dispersion Spectrograph 2 (FORS2) mounted the
volume-phased holographic grism GRIS\_1400V$+$18 with 1400 $\rm
grooves\;mm^{-1}$ and the $0\farcs7\,\times\,6\farcm8$ slit.
The detector was a mosaic of 2 MIT/LL CCDs. Each CCD had
$2048\times2068$ pixels of $15\,\times\,15$ $\mu$m$^2$. The wavelength
range from 4560 to 5860 \AA\ was covered with a reciprocal dispersion
of 0.645 \AA\ pixel$^{-1}$ and a spatial scale of 0.250 arcsec
pixel$^{-1}$ after a $2\times2$ pixel binning.
We obtained spectra with the slit at $\rm PA=7\fdg9$ crossing the
galaxy center ($3\times45$ minutes in run 1) and offset by $5\farcs0$
eastward ($4\times30$ minutes in run 3). We also obtained spectra with
the slit at $\rm PA=6\fdg9$ and shifted by $5\farcs0$ eastward
($2\times40$ minutes in run 2; $2\times40$ minutes in run 3) and
westward ($4\times40$ minutes in run 3) with respect to the galaxy
nucleus. Two different PAs were chosen for the slits to test for the
sensitivity of the TW measurement to the $\sim 1^\circ$ uncertainty in
$\rm PA_{\it disk}$.
Comparison lamp exposures obtained for each observing night ensured 
accurate wavelength calibrations. Spectra of G and K giant stars 
served as kinematical templates. The average seeing FWHM was 
$1\farcs2$ in run 1, $0\farcs8$ in run 2, and $0\farcs9$ in run 3. 
Using standard {\tt MIDAS}\footnote{{\tt MIDAS} is developed and
maintained by the European Southern Observatory.} routines, all the
spectra were bias subtracted, flat-field corrected, cleaned of cosmic
rays, corrected for bad pixels, and wavelength calibrated as in
Debattista et al. (2002).  The accuracy of the wavelength rebinning
($\approx1$ \kms) was checked by measuring wavelengths of the
brightest night-sky emission lines. The instrumental resolution was
$1.45\pm0.01$ \AA\ (FWHM) corresponding to $\sigma_{\it inst} = 35$
\kms\ at 5170 \AA .
The spectra obtained in the same run along the same position were
co-added using the center of the stellar continuum as reference. In
the resulting spectra the sky contribution was determined by
interpolating along the outermost $\approx30''$ at the edges of the
slit and then subtracted.

\section{Pattern speed of the bar}      
\label{sec:pattern} 
      
To measure the pattern speed of the bar, $\om$, we used the TW
method. For slits parallel to the disk major-axis, this relates the
luminosity-weighted mean position, $\pin$, to the luminosity-weighted
mean velocity $\kin = \pin \om \sin i$ where $i$ is the galaxy
inclination.
 
To measure $\kin$ for each slit, we first collapsed the
two-dimensional spectrum along its spatial direction between $-90''$
and $+90''$ in the wavelength range between 5220 and 5540 \AA,
obtaining a one-dimensional spectrum.  The value of $\kin$ was then
derived by fitting the resulting spectrum with the convolution of the
spectrum of the G0III star SAO119458 and a Gaussian line-of-sight
velocity profile by means of the Fourier Correlation Quotient method
(Bender 1990, hereafter FCQ) as done in Aguerri et al. (2003,
hereafter ADC03). We estimated uncertainties by Monte Carlo
simulations with photon, read-out and sky noise. As a precision check
we analyzed separately the two eastward-offset spectra ($Y=+5\farcs0$)
which were obtained with the slit at $\rm PA=6\fdg9$ in run 2 and 3,
respectively. The two resulting estimates of $\kin$ are in agreement
within the errors (Fig. 2).
To compute $\pin$ for each slit, we extracted the luminosity profile
from the $R-$band image along the position of the slit after
convolving the image to the seeing of the spectrum. The $R-$band
profile matches very well that obtained by collapsing the spectrum
along the wavelength direction, supporting that the slit was placed as
intended. We used the $R-$band profiles to compute $\pin$ because they
are less noisy than those extracted from the spectra, particularly at
large radii.  By comparing the slits at $\rm PA=6\fdg9$ and at $\rm
PA=7\fdg9$ we also verified that the error introduced by the $\sim
1^\circ$ uncertainty in disk PA is small for both $\pin$ and $\kin$.
We obtained $\om$ by fitting a straight line to the values of $\pin$
and $\kin$ to the data for slits at $Y=\pm5\farcs0$ and $\rm
PA=6\fdg9$ and at $Y=0''$ and $\rm PA=7\fdg9$ (Fig. 2). Although the
$Y=0''$ slit was obtained at different PA, this slit constrains only
the zero point (i.e., systemic) velocity and any slit passing through
the center can be used. This gives $\om \sin{i} = 5.56
\pm 1.39$ \kms\ arcsec$^{-1}$ ($70.8 \pm 17.7 $ \kms\ kpc$^{-1}$).
 
We used the FCQ to measure the line-of-sight velocity curve and
velocity dispersion profile of the stellar component along the major
axis (Fig. 3). Our measurements at $\rm PA=6\fdg9$ are in agreement
within errors with those by Simien \& Prugniel (2002) at $\rm
PA=6^\circ$.
We derived the circular velocity in the disk region, $V_c = 94\pm19$
\kms, after a standard correction for the asymmetric drift as in
ADC03.  This value is in agreement with the circular velocity derived
from the Tully-Fisher relation calculated in the $R-$band by Courteau
(1997). Thus the corotation radius of the bar is $\lag = V_c/\om =
12\farcs7_{-2.9}^{+4.3}$ and the ratio of the corotation radius to the
bar semi-major axis $\vpd \equiv \lag/\len = 0.6_{-0.4}^{+1.2}$. The
error intervals on $\lag$ and $\vpd$ are $99\%$ confidence level and
were measured with Monte Carlo simulations as in ADC03.
By excluding the possibility that $\lag/\len < 1.0$ as this is thought
to be unphysical (Contopoulos 1980) we estimated that the probability
that the bar ends close to its corotation radius (i.e., $1.0 \leq
\lag/\len \leq 1.4$) is $72\%$. This was calculated via Monte Carlo
simulations by assuming a uniform distribution of $a_B$ and $V_c$ and
Gaussian distribution of $\om$ within their error ranges.

\section{Discussion} 
\label{sec:discussion} 

By measuring the bar pattern speed of NGC~4431 we have demonstrated
that the TW method is feasible using stars as tracer for dwarf
galaxies.
So far, this technique has been applied only to bright barred galaxies
with the result that all the measured bars are consistent with being
rapidly rotating (see Corsini 2004 for a review).
For NGC 4431 we find a bar suggestive of being fast at $72\%$
probability, although the measurement uncertainties in this galaxy
preclude a stronger statement.

As discussed by Debattista (2003) and Debattista
\& Williams (2004), the TW measurement of $\om$ is particularly
sensitive to errors in the PA$_{\it disk}$.  
However, this is unlike to be an issue for NGC~4431. In fact, the
values of $\pin$ and $\kin$ measured in the offset position at
$Y=+5\farcs0$ from the spectrum at $\rm PA=7\fdg9$ are in agreement
within the errors with those from the two spectra at $\rm PA=6\fdg9$.
So, the fact that the value of the bar speed parameter of NGC~4431 is
nominally less than unity cannot be attributed to the uncertainty on
PA$_{\it disk}$ but to the scatter in the values of $\pin$ and $\kin$.

A fast bar in NGC~4431 would suggest a common formation mechanism of
the bar both in bright and dwarf galaxies. If disks were previously
stabilized by massive DM halos, we exclude that these bars were
produced by tidal interactions because they would be slowly rotating
(Noguchi 1999).
But, this is not the case even for galaxies, like NGC~4431 and the SB0
NGC~1023 (Debattista et al. 2002) which show signs of weak interaction
with a close companion, without being significantly perturbed. This
finding allows us to conclude that there is no difference between TW
measurements in isolated or mildly interacting barred galaxies (ADC03;
Corsini et al. 2003).

According to high-resolution $N$-body simulations of bars in
cosmologically motivated, CDM halos (see Debattista 2006 for a review)
a fast bar would imply that NGC~4431 has a massive disk and does not
reside inside a centrally-concentrated DM halo.
This is in agreement with other studies of dwarf galaxies (Swaters
1999; de Blok et al. 2001; McGaugh et al. 2001; de Blok \& Bosma 2002;
Marchesini et al. 2002; Swaters et al. 2003; Spekkens et al. 2005) who
analyzed the ionized-gas rotation curves of up to about $200$
objects. By assuming minimal disks and spherical symmetry most of the
data are better described by DM halos with an inner density core
rather than the cusp predicted by CDM.

However, there is no consensus whether these results based on gas
dynamics represent an actual problem for the CDM paradigm or whether
they could be reconciled with it by taking a variety of observational
uncertainties (e.g., slit positioning, spatial resolution), analysis
problems (e.g., binning and folding of the rotation curves), and
modelling assumptions (e.g., non-circular and off-planar gas motions)
into account (de Blok et al. 2003; Swaters et al. 2003; Spekkens et
al. 2005). Therefore, we argue that the ongoing debate will benefit
from stellar dynamics to constrain the inner mass distribution of DM
with the direct measurement of $\om$ in a sample of dwarf barred
galaxies by means of integral-field spectroscopy.

\bigskip 
\noindent 
{\bf Acknowledgments.}  
 
\noindent 
EMC and AP receive support from the grant PRIN2005/32 by
Istituto Nazionale di Astrofisica (INAF) and from the grant
CPDA068415/06 by the Padua University.
JALA has been funded by the Spanish DGES, grant AYA-2004-08260-C03-01.
VPD is supported by a Brooks Prize Fellowship in Astrophysics at the
University of Washington and receives partial support from NSF ITR
grant PHY-0205413.
FDB acknowledges support from the LTSA grant NAG5-13063 and from the
NSF grant AST-0607748.
VPD and EMC acknowledge the Instituto de Astrof\'{\i}sica de Canarias
for hospitality while this paper was in progress.

\clearpage   

\begin{figure}
\plotone{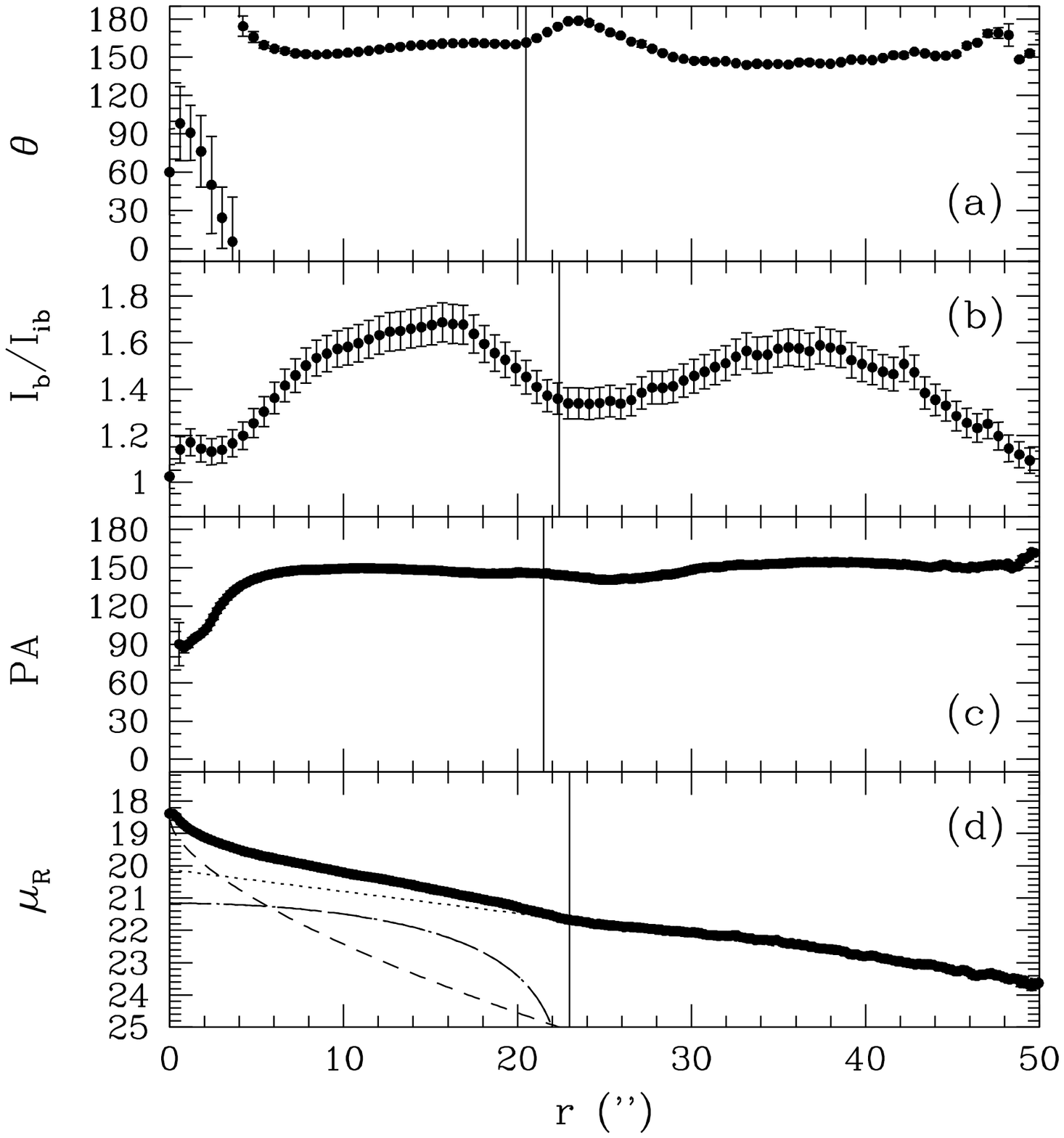}
\caption{Bar semi-major axis length of NGC 4431. {\it (a)\/}  
   Phase angle of $m=2$ Fourier component. {\it (b)\/} Bar/interbar
   intensity ratio. {\it (c)\/} PA of the deprojected isophotal
   ellipses. {\it (d)\/} Surface brightness decomposition in bulge
   (dashed line), disk (dotted line), and bar (dashed-dotted line). The
   vertical lines show the value of $\len$ obtained with each model.}
\end{figure} 

\clearpage

\begin{figure}
\plotone{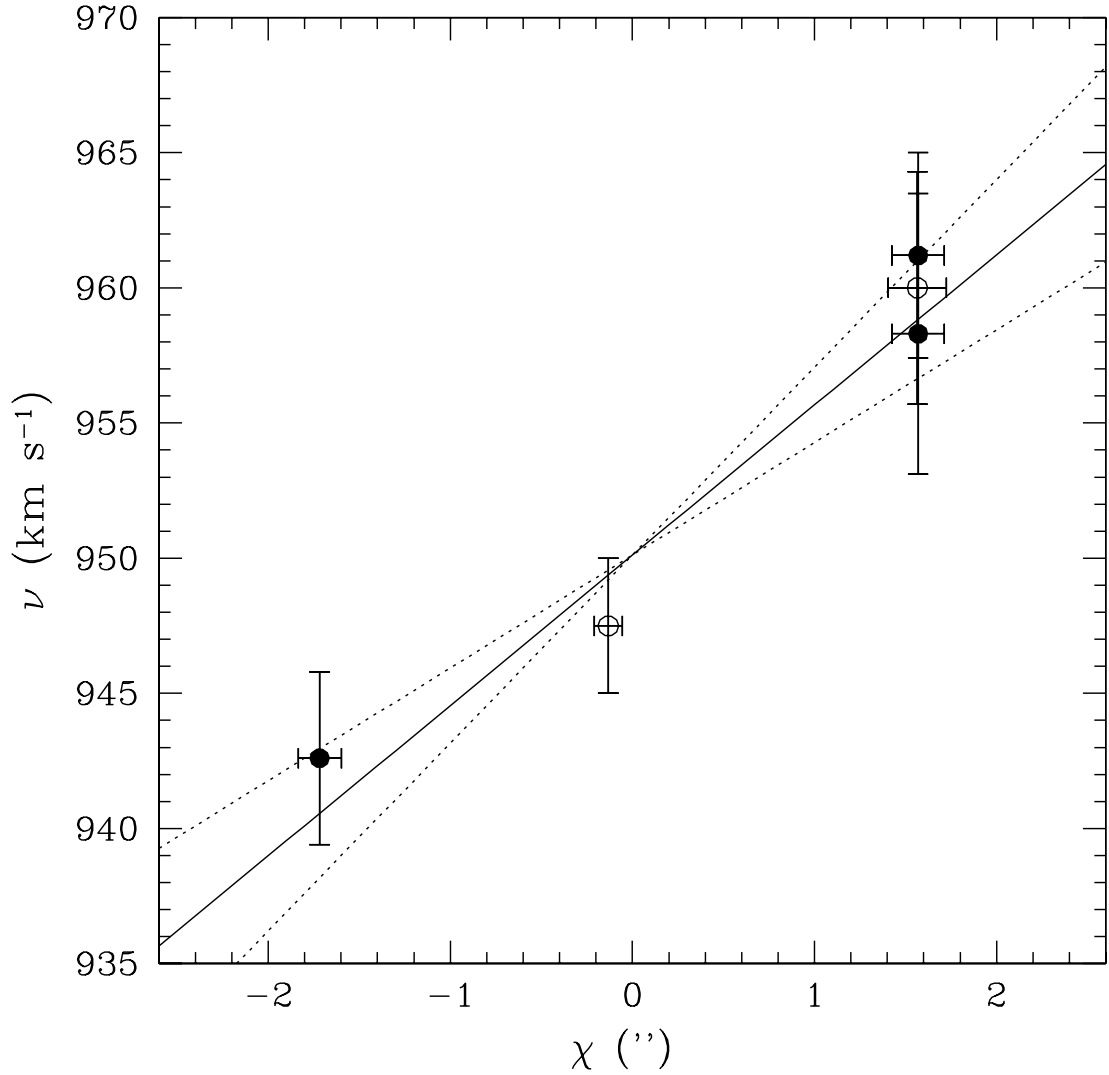}
\caption{Pattern speed of the bar in NGC~4431.  The kinematic
    integrals ${\cal V}$ are plotted as a function of the photometric
    integrals ${\cal X}$. The best fitting straight line has a slope
    $\om \sin{i} = 5.56 \pm 1.39$ \kms\ arcsec$^{-1}$. Open circles
    correspond to slits at $\rm PA=7\fdg9$ ($Y=0''$ and
    $Y=+5\farcs0$), while filled circles are for slits at $\rm
    PA=6\fdg9$ ($Y = \pm5\farcs0$).}
\end{figure} 

\clearpage

\begin{figure}
\plotone{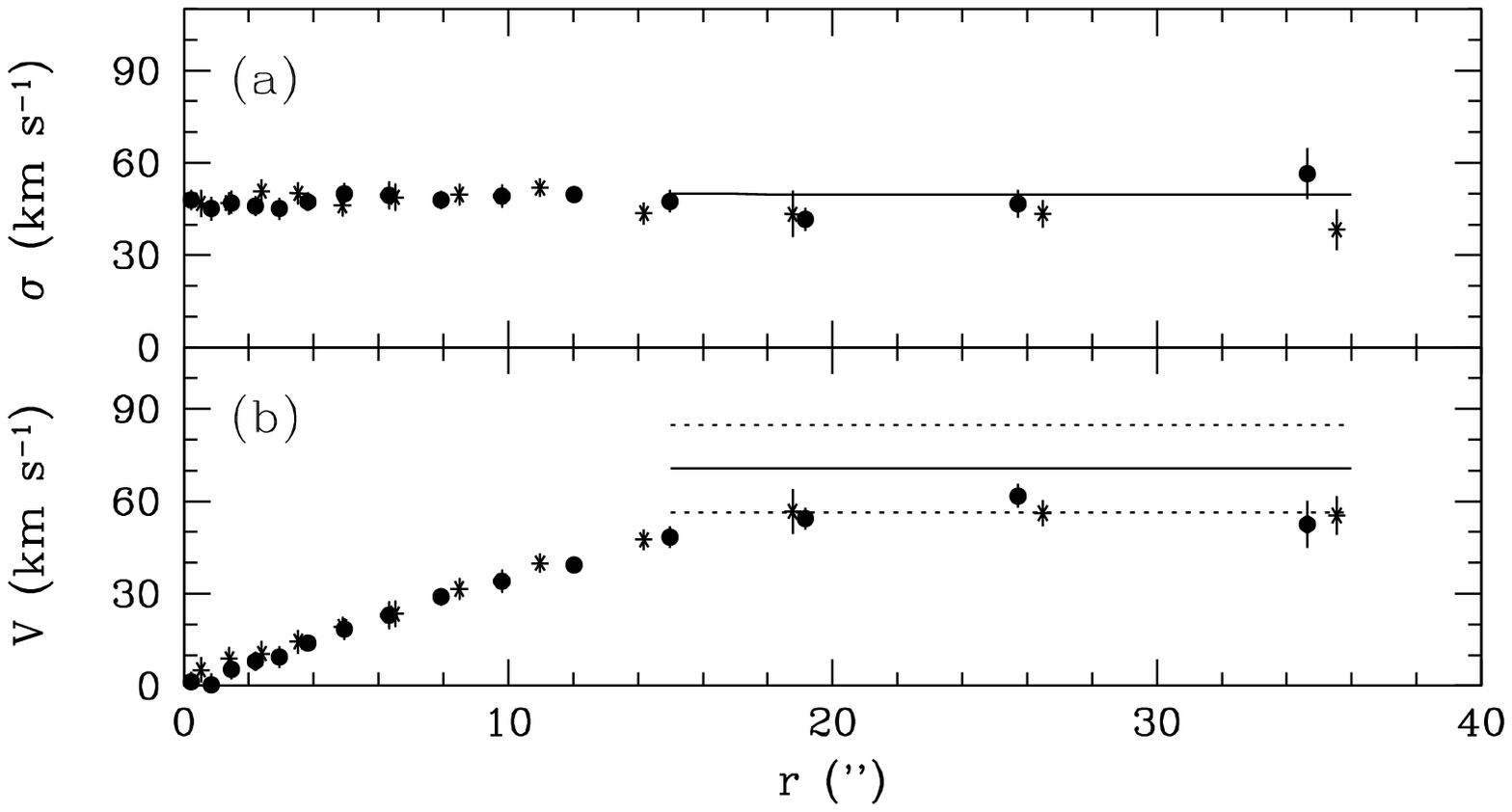}
\caption{{\it (a)\/} The major-axis radial profile of the stellar  
    line-of-sight velocity dispersion fitted with an exponential  
    profile at $r\geq15''$ (solid line).  
    {\it (b)\/} The major-axis radial profile of the stellar line-of-sight  
    velocity (after subtracting the systemic velocity $V_{\it  
      sys\/}=948\pm3$ \kms) and the $V_c \sin{i}$ curve (solid line)  
    with errors (dotted lines) obtained by applying the asymmetric drift for  
    $r\geq15''$ as in ADC03.  
    In {\it (a)\/} and {\it (b)\/} the measured profiles are folded 
    around the center, with filled circles and asterisks referring to 
    the N (receding) and S (approaching) sides, respectively.} 
\end{figure} 


\bigskip 
\noindent 
 
\clearpage
 
\begin{thebibliography}{} 
 
 
\bibitem[]{422} Aguerri J. A. L., Mu\~noz-Tu\~n\'on C., Varela A. M.,  
             \& Prieto M. 2000, \aap, 361, 841 

\bibitem[]{425} Aguerri, J.\ A.\ L., Debattista, V.\ P., Corsini, E.\ M.\ 
             2003, \mnras, 338, 465 (ADC03) 
    
\bibitem[]{428} Athanassoula, E. 2003, \mnras, 341, 1179 
  
\bibitem[]{430} Barazza, F.\ D., Binggeli, B., \& Jerjen, H.\ 2002, \aap,  
             391, 823 

\bibitem[]{433} Barazza, F.\ D., Binggeli, B., \& Jerjen, H.\ 2003, \aap,  
             407, 121 

\bibitem[]{436} Bender, R. 1990, \aap, 229, 441 

\bibitem[]{438} Binggeli, B., Sandage, A., \& Tammann, G.~A.\ 1985,  
             \aj, 90, 1681  

\bibitem[]{441} Bosma, A. 1999, in Galaxy Dynamics, eds. D. Merritt,
             J. A. Sellwood, \& M. Valluri, ASP Conf. Ser. 182, 
             (San Francisco: ASP), p. 339

\bibitem[]{900} Bureau, M., Freeman, K.~C., Pfitzner, D.~W., \& Meurer, G.~R.\ 
             1999, \aj, 118, 215

\bibitem[]{445} Corsini, E.\ M., 2004, in Baryons in Dark Matter  
             Halos, eds. R. Dettmar, U. Klein, \& P. Salucci, 
             (Trieste: SISSA) 
             p. 49.1 
 
\bibitem[]{450} Corsini, E.~M., et al.\ 1999, \aap, 342, 671  
 
\bibitem[]{901} Corsini, E.\ E., Debattista, V.\ P., Aguerri, J.\ A.\ L.\
             2003, \apj, 599, L29
    
\bibitem[]{452} Courteau, S.\ 1997, \aj, 114, 2402 

\bibitem[]{454} Contopoulos, G.\ 1980, \aap, 81, 198
 
\bibitem[]{456} Debattista, V.~P., 2003, \mnras, 342, 1194 

\bibitem[]{458} Debattista, V.~P., 2006, in New Horizons in Astronomy,
             eds. S. Kannappan et al., ASP Conf. Ser. 352 (San Francisco: 
             ASP), p. 161
 
\bibitem[]{462} Debattista, V.~P., \& Sellwood, J.~A., 1998, \apj, 493, L5 
 
\bibitem[]{464} Debattista, V.~P., \& Sellwood, J.~A., 2000, \apj, 543, 704 

\bibitem[]{466} Debattista, V.~P., \& Williams, T.~B., 2004, \apj, 605, 714 

\bibitem[]{468} Debattista, V.~P., Corsini, E.~M., \& Aguerri, J.~A.~L.  
             2002, \mnras, 332, 65 

\bibitem[]{471} de Blok, W.~J.~G., McGaugh, S., \& Rubin, V.~C.\ 2001, \aj, 
             122, 2396

\bibitem[]{474} de Blok, W.~J.~G., \& Bosma, A.\ 2002, \aap, 385, 816  
 
\bibitem[]{476} de Vaucouleurs, G., de Vaucouleurs, A., Corwin Jr., H. G., 
             Buta, R. J., Paturel, G., \& Fouqu\`e, P. 1991, 
             Third Reference Catalogue of Bright Galaxies (New York: 
             Springer) (RC3) 
 
\bibitem[]{481} Eskridge, P., et al.\ 2000, \aj, 119, 563 
 
\bibitem[]{483} Hayashi, E., \& Navarro, J. F. 2006, \mnras, 373, 1117

\bibitem[]{485} Knapen J. H., Shlosman, I., \& Peletier, R. F. 2000, ApJ, 
             529, 93

\bibitem[]{488} Jerjen, H., Binggeli, B., \& Barazza, F.\ D.\  
             2004, \aj, 127, 771 

\bibitem[]{491} Marinova, I., \& Jogee, S. 2006, ApJ, in press (astro-ph/0608039)

\bibitem[]{493} Marchesini, D., D'Onghia, E., Chincarini, G., Firmani, C., 
             Conconi, P., Molinari, E., \& Zacchei, A.\ 2002, \apj, 
             575, 801 

\bibitem[]{497} McGaugh, S., Rubin, V.~C., \& de Blok, W.~J.~G. 2001, \aj, 
             122, 2381

\bibitem[]{500} Moore, B., Governato, F., Quinn, T., Stadel, J., 
             \& Lake, G.\ 1998, \apjl, 499, L5 

\bibitem[]{503} Moore, B., Quinn, T., Governato, F., Stadel, J., 
             \& Lake, G.\ 1999, \mnras, 310, 1147 
 
\bibitem[]{506} Navarro, J.~F., Frenk, C.~S., \& White, S.~D.~M.\ 
             1996, \apj, 462, 563 

\bibitem[]{509} Navarro, J.~F., Frenk, C.~S., \& White, S.~D.~M.\ 
             1997, \apj, 490, 493 

\bibitem[]{512} Noguchi, M.\ 1999, \apj, 514, 77 

\bibitem[]{514} O'Neill, J.~K., Dubinski, J.\ 2003, \mnras, 346, 251 
 
\bibitem[]{516} Pedraz, S., Gorgas, J., Cardiel, N., S\'anchez-Bl\'azquez, 
             P., \& Guzm\'an, R.\ 2002, \mnras, 332, L59 
 
\bibitem[]{519} Prieto, M., Aguerri, J.~A.~L., Varela, A.~M., \& 
             Mu\~ noz-Tu\~n\'on, C. 2001, \aap, 367, 405 
 
\bibitem[]{522} Sellwood, J.~A., \& Debattista, V.~P., 2006, \apj,  
             639, 868
 
\bibitem[]{525} Simien, F., \& Prugniel, P.\ 2002, \aap, 384, 371 

\bibitem[]{527} Spekkens, K., Giovanelli, R., \& Haynes, M. P. H. 2005, 
             \aj, 129, 2119

\bibitem[]{530} Swaters, R.~A., 1999, PhD thesis, University of 
             Groningen
 
\bibitem[]{533} Swaters, R.~A., Madore, B.~F., van den Bosch,  
             F.~C., \& Balcells, M.\ 2003, \apj, 583, 732 
   
\bibitem[]{536} Tremaine, S., \& Weinberg, M.~D., 1984, \apj, 282, L5 
 
\bibitem[]{538} Valenzuela, O., Rhee, G., Klypin, A., Governato, F., Stinson, 
             G., Quinn, T.; Wadsley, J.\ 2005, \apj, submitted 
             (astro-ph/0509644)

\bibitem[]{542} van den Bosch, F.~C., \& Swaters, R.~A.\ 2001,  
             \mnras, 325, 1017 

\bibitem[]{545} Weinberg, M.~D., 1985, \mnras, 213, 451 

\end{thebibliography}
\end{document}